\documentclass [twocolumn,epsfig,prl,10pt,floatfix,amsmath,amssym]{revtex4}
\usepackage[dvips]{epsfig}

\usepackage{float}

\newcommand{\xv}{{\bf x}}

\newcommand{\uv}{{\bf u}}
\newcommand{\Dv}{{\bf \Delta}}

\newcommand{\tv}{\hat{{\bf t}}}

\newcommand{\grad}{{\bf \nabla}}

\begin{document}

\title{Topological Defects in Twisted Bundles of Two-Dimensionally Ordered Filaments}
\author{Gregory M. Grason}
\affiliation{Department of Polymer Science and Engineering, University of Massachusetts, Amherst, MA 01003}

\begin{abstract}

Twisted assemblies of filaments in ropes, cables and bundles are essential structural elements in wide use in macroscopic materials as well as within the cells and tissues of living organisms.  We develop the unique, non-linear elastic properties of twisted filament bundles that derive from generic properties of two-dimensional line-ordered materials.  Continuum elasticity reveals a formal equivalence between the elastic stresses induced by bundle twist and those induced by the positive curvature in thin, elastic sheets.  These geometrically-induced stresses can be screened by 5-fold disclination defects in lattice packing, and we predict a discrete spectrum elastic energy groundstates associated with integer numbers of disclinations in cylindrical bundles.  Finally, we show that elastic-energy groundstates are extremely sensitive to defect position in the cross-section, with off-center disclinations driving the entire bundle to buckle, adopting globally writhing configurations.  
\end{abstract}
\date{\today}

\maketitle

From the textile fibers in our clothing to the steel cables that suspend the Brooklyn bridge, twisted filament assemblies are familiar structural elements of everyday life.  This deceptively simple geometric motif, a dense packing of helical filaments with a constant pitch and varying radius, underlies several unique mechanical properties of ropes and fibers.  The twist of a rope or yarn transmits tension to lateral compression, binding together multiple short strands into single cohesive element of arbitrary length~\cite{hearle}.  For the case of wire ropes, composed of continuous filaments, helical pre-twist allows one to build a material with a high tensile strength while maintaining remarkable flexibility from an assembly of relatively inextensible elements~\cite{costello}.  It is not surprising that Nature has evolved many ways to take advantage of the mechanical fidelity of ropes, at much smaller length scales, by way of super-molecular assemblies of protein filaments in living organisms.  Bundles of cytoskeletal filaments, f-actin or microtubules, are implicated in a variety of cellular processes in eukaryotic life, from cell division to mechanosensing~\cite{theriot}.  Fibers of extra-cellular proteins, like collagen, are ubiquitous mechanical components of plant and animal tissue~\cite{fratzl}.  Because biological filaments are helically-structured molecules, inter-molecular forces between densely-packed filaments generate intrinsic torques in bundle assemblies.  Indeed, there is clear evidence of spontaneous twisting of a variety biological filament bundle types~\cite{weisel, wess, ikawa}, pointing to yet another function of twist in these self-organized bundles.  The frustration between two-dimensional packing and inter-filament twist generically gives rise to a mechanism of {\it self-limited} radial bundle growth~\cite{turner, grason_bruinsma, grason}, suggesting that spontaneous bundle twist plays a necessary role in shaping the structure and mechanics of biological filament assemblies.  

While mechanical considerations of ropes and fibers date back to as early as Galileo, very little is known about the role that the cross-sectional ordering plays in determining the global elastic properties of twisted bundles.  In this Letter, we describe a continuum approach to the unique, non-linear elasticity of two-dimensionally (2D) ordered filament bundles, revealing a fundamental coupling between the presence and location of {\it topological defects} in the lattice packing and the global twist of the bundle.  We discover a precise equivalence between the in-plane stresses created by twisting a bundle and the long-range stresses introduced into a 2D elastic sheet by forcing it to adopt a spherically curved geometry.  Due to the accumulation of elastic stresses, we show that a sufficiently large and twisted bundle becomes energetically unstable to the incorporation of finite number of 5-fold disclination defects in its cross-section.  We predict a spectrum of discrete low elastic energy states of twisted bundles, associated with a integer numbers of disclinations as a function of increasing bundle twist.  Finally, we show that the global geometry of the low-energy states of bundles is extremely sensitive to the location of disclinations in the cross-section.  Moving the position of disclination away from the center of the bundle leads to a complex class of structures that buckle, writhing out of the plane, ultimately folding the entire bundle into a compact helical configuration.

Consider a bundle of semi-flexible filaments of infinite length that is organized into a regular array perpendicular to mean filament orientation along the $\hat{z}$ direction.  Interactions between filaments favor a constant neighbor spacing, $d_0$, much smaller than the radius of the bundle, $R$, which is approximately cylindrical in shape.  The zero energy state of these interactions is a hexagonal packing of straight filaments.   Deviations from the uniform spacing as well as filament bending are described by the energy~\cite{grason},
\begin{equation}
\label{eq: en}
E = \frac{1}{2} \int d^3 x \Big\{ \lambda (u_{ii})^2 + 2 \mu u_{ij} u_{ij} + K_b [(\tv \cdot \grad) \tv]^2\Big\} .
\end{equation}
Here, $\lambda$ and $\mu$ are elastic constants that penalize compressive and shear distortions of the array and $u_{ij}$ is elastic strain tensor of the 2D filament bundle.  This strain is constructed from displacements of filament positions, $\uv(\xv)$, in the $xy$ plane, and the projection of the local filament orientation, $\tv(\xv)$, into the $xy$ plane,
\begin{equation}
\label{eq: u}
u_{ij} = \frac{1}{2}(\partial_i u_j+ \partial_j u_i - t_i t_j) .
\end{equation}
The non-linear contribution to strain from filament tilt is a direct consequence of the invariance of the elastic energy under rotation about in axis in $xy$ plane~\cite{selinger_bruinsma}, and can be intuitively understood to measure distances between neighbor filaments {\it perpendicular to the filament tangent}, that is, the distance of closest approach.  The final term in eq. (\ref{eq: en}) describes the bend elasticity of filaments, with $K_b$ proportional to the stiffness of filaments.  

To study low elastic energy configurations of twisted filament bundles whose cross section contains an arbitrary distribution of disclination defects, we focus on configurations for which the shear and compressive contributions to eq. (\ref{eq: en}) are constant along filaments.  In theses structures, $\uv(\xv)$ and $\tv_\perp(\xv)$ differ along $z$ only by uniform rotations around an axis in the $xy$ plane, and at some reference plane ($z=0$), the in-plane tilt field is described by $\tv_\perp (\xv) = \Dv + \Omega (\hat{z} \times \xv) $ where $\xv = 0$ is the center of the bundle, $\Dv$ is a constant tilt in the plane, and $\Omega$ is the rate of torsion.  Initially, we restrict our analysis to $\Dv = 0$, in which filaments wind around the geometrical center of the bundle.  
 
For fixed $\Omega$, configurations are determined by $\partial_i \sigma_{ij} = 0$ subject to stress-free boundary conditions at the surface of the bundle.  As we have reduced the problem to solving for the stress $\sigma_{ij} = \delta E/\delta u_{ij}$ in a single 2D elastic plane, it is convenient to determine the stress tensor in terms an Airy stress function, $\chi$.  By construction, $\sigma_{ij} = \epsilon_{ik} \epsilon_{j \ell} \partial_k \partial_\ell \chi$, is divergenceless, and the Airy stress is determined by from ``compatibility relation" which relates $\chi$ to the definition of elastic strain, eq. (\ref{eq: u}),
\begin{multline}
\label{eq: compat}
K_0^{-1} \grad_\perp^4 \chi = \sum_\alpha s_ \alpha \delta (\xv-\xv_\alpha) \\ -\frac{1}{2} \grad_\perp \times \big[(\grad_\perp \times \tv_\perp) \tv_\perp - (\tv_\perp \times \grad_\perp) \tv_\perp \big],
\end{multline}
where $K_0 = 4 \mu( \lambda +  \mu)/(\lambda + 2 \mu)$ and $\grad_\perp \times \tv_\perp = \epsilon_{ij} \partial_i t_j$ is the 2D curl.  Here, $s_ \alpha = \pm 2 \pi/6$ denotes the topological charge of a disclination at position, $\xv_ \alpha $, in the cross-section of the bundle where the local bond angle, $\theta(\xv) = \epsilon_{ij} \partial_i u_j/2$ rotates around singular points.  Disclinations represent lattice sites of 5- (positive) and 7-fold (negative) symmetry in the otherwise 6-fold symmetric packing.  The second line of eq. (\ref{eq: compat}) describes the non-linear contribution from in-plane tilts to the strain in the bundle, deriving from the fact that in 2D ordered chains, twisted configurations of tangents necessarily introduce gradients in nearest neighbor orientation, lying in a perpendicular plane~\cite{kamien}.  For the case of a twisted bundle where $\grad_\perp \times \tv_\perp  = \pm 2 \Omega+{\cal O} [(\Omega r)^3]$, the second line reduces to $- K_{eff} =-\epsilon_{ij} \epsilon_{k \ell} \partial_i \partial_{k} t_j t_\ell / 2=  - 3 \Omega^2$, a uniform and negative source of Airy stress in the bundle.

\begin{figure}
\center \epsfig{file=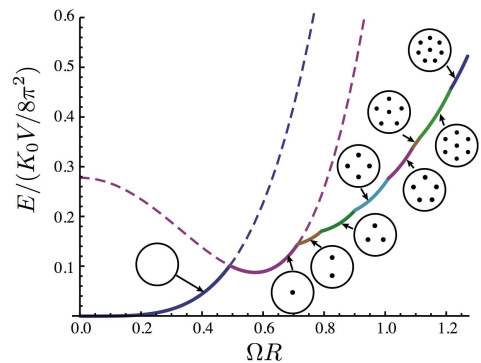, width=2.45in}\caption{Solid curves show elastic energy of 2D bundle groundstates a function of twist times bundle radius, $\Omega R$, for $K_b =0$.  The cartoons depict the energy minimizing configurations of multiple 5-disclinations.  The dashed lines show the metastable branches of the disclination-free and single-disclination bundle energy.} 
\label{fig: figure1}
\end{figure}

An analogous compatibility equation describes the non-linear elasticity of 2D crystalline membranes, known as the F\"oppl-van K\'arm\'an theory of thin plates~\cite{landau}.  In that case, the coupling between in-plane strain and out-of-plane bending generates a source of the Airy stress, $-K$, where $K$ is the Gaussian curvature of the surface.  Hence, we find the surprising result that twist of filament bundle generates precisely the same source of stress as stretching a 2D elastic sheet over a positively curved, spherical surface of radius $R_{eff} = (\sqrt{3} \Omega)^{-1}$.  Seung and Nelson showed that the far-field stresses created by Gaussian curvature in 2D crystalline membranes are precisely of the form to be ``screened" by the presence of point-like disclination defects in the membrane~\cite{nelson_seung}.  For the case of 2D filament bundles, eq. (\ref{eq: compat}) shows that positively charged, 5-fold disclinations are able to partially screen the elastic stresses of twisted bundles.

The elastic energy of twisted bundles with an arbitrary distribution of disclinations can be computed by solving eq. (\ref{eq: compat}) for $\chi$ subject to vanishing normal stress at $r=R$ by multipole expansion~\cite{brown}.  These expansions can be summed to compute the effective theory for the elastic energy,
\begin{multline}
\label{eq: Eeff}
\frac{E}{V K_0} = \sum_\alpha \frac{s_\alpha^2}{32 \pi^2} \Big(1-\frac{\rho_\alpha^2}{R^2}\Big)^2 - \frac{3 (\Omega R)^2 }{2} \sum_\alpha \frac{s_\alpha}{32 \pi}  \Big(1-\frac{\rho_\alpha^2}{R^2}\Big)^2 \\ 
+ \frac{3(\Omega R)^4}{128}\Big(1+\frac{3}{32 \gamma} \Big) + \frac{1}{2} \sum_{\alpha \neq \beta} s_\alpha V_{int}(\xv_\alpha,\xv_\beta) s_\beta ,
\end{multline}
where $V$ is the volume of the bundle and $\rho_\alpha$ is the radial distance the $\alpha$th disclination.  The first term describes the self-energy of a disclination in the bundle, which notably, has the same radial dependence as the interaction between the disclination and twisted-induced stress described the second term.  The third term describes the twist dependence of elastic energy in the defect-free state, where $\gamma \equiv K_0 R^2/K_b$, that measures the relative importance in-plane elastic energy to bend elasticity.  The final term gives the effective interaction between disclination pairs in cylindrical crystals,
\begin{multline}
V_{int}(\xv_\alpha,\xv_\beta)= \frac{1}{16 \pi^2} \Big(1-\frac{\rho_\alpha^2}{R^2}\Big)\Big(1-\frac{\rho_\beta^2}{R^2}\Big)\\  + \frac{|\xv_\alpha-\xv_\beta|^2}{8 \pi^2 R^2} \ln \bigg(\frac{|\xv_\alpha-\xv_\beta|}{|\xv'-{\bf R}|}\bigg),
\end{multline}
where $|\xv'-{\bf R}|^2 = R^2+\rho_\alpha^2 \rho_\beta^2/R^2 - 2\rho_ \alpha \rho_\beta \cos \theta$, with $\theta$ the angular separation between disclinations $\alpha$ and $\beta$.  The minimum-energy configurations of bundles with 5-fold ($s_\alpha = +2\pi/6$) disclinations are shown in Figure \ref{fig: figure1} for $\Omega R \leq 1.25$ and for negligible bending energy, $\gamma^{-1} =0$ .  For these results we find that the elastic energy of a filament bundle is unstable to the incorporation a finite number of disclinations when the twist of a bundle exceeds $|\Omega_c R| = \sqrt{2/9} \simeq 0.471$, at which point the lowest energy is accomplished by placing a single 5-fold defect at the center.  Upon further twist, this single defect structure gives way to lower energy states with 2, 3, 4, 5... disclinations.  Evidently, along with crystals on curved surfaces, twisted bundles belong to the unusual class of frustrated materials in which defects are necessary components of groundstate configurations.  We note that although 7-fold disclinations are not necessary to screen the twist-induced stresses in bundle, we expect these defects to appear in low-energy configurations when $R\gg d_0$.  Just as 5-fold disclinations on large spherical crystals~\cite{bowick} are unstable to breaking up into extended strings of alternating 5- and 7-fold defects of a net positive charge, so-called grain boundary scars, we anticipate similar extended defect configurations in the elastic-energy ground states of large twisted bundles.  

The elastic energy in eq. (\ref{eq: Eeff}) demonstrates an important and previously unknown coupling between the in-plane stresses created by lattice defects and bundles twist:  5-fold disclinations screen the twisted-induced stresses and {\it visa versa}.  Hence, the elastic energy of bundles with an excess of 5-fold disclinations (relative to 7-fold) can always be lowered by twisting a finite amount.    The right- or left-handed spontaneous twist of a bundle is the analog of the upward or downward buckle of a crystalline membrane induced by the presence of 5-fold defects in the hexagonal order~\cite{nelson_seung}.  The degree of ``torsional buckling" is sensitive to the relative costs of filament bending and in-plane elastic energies.  We find that optimal degree of twist, $\Omega_1$,  for a bundle with a single centered, 5-fold disclination has the form $(\Omega_1 R)^2= (3+32/\gamma)^{-1}$.  For small bundles of rigid filaments, $\gamma \to 0$, the pitch of the bundle sensitive to the bending the elastic moduli and insensitive to size, $2 \pi /\Omega_1 \propto \sqrt{K_0/K_b}$.  While for bundles much larger than this characteristic size, $\gamma \to \infty$, the tilt angle of the outer filament is predicted by the continuum theory to achieve a universal value, $\Omega_1 R \to \sqrt{1/3} \simeq 0.577$.

\begin{figure}
\center \epsfig{file=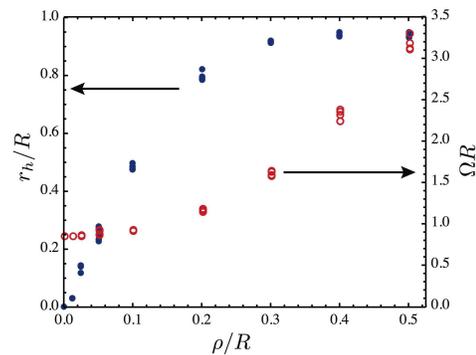, width=2.45in}\caption{The groundstate geometry of elastic energy of eq. (\ref{eq: num}), for $R=10 d_0$ bundles possessing a single disclination at radius $\rho$.  The filled, blue circles show the helical radius of the geometrical center of the bundle and the open, red circles show the rate of torsion around the $z$ axis.  The data for each $\rho$ correspond to disclinations at 6 azimuthal angles:  $0, \pm 2\pi/6, \pm 4\pi/6$ and $\pi$.} 
\label{fig: figure2}
\end{figure}

We now consider filament bundles for which $\Dv \neq 0$, with successive vertical layers are rotated and shifted by a constant amount in the $xy$ plane.  From the continuum theory we compute the mean torque on the filaments due to the presence of twist- and defect-induced stress,
\begin{multline}
\label{eq: torque}
\frac{\partial E}{\partial \Delta_i} = -\int d^3 x ~ \sigma_{ij} t_j  = - \frac{V K_0 \Omega }{8 \pi} \\ \times \sum_\alpha s_\alpha  \Big[ 1 - \frac{\rho_\alpha^2}{R^2} - 2  \frac{\rho_ \alpha ^2}{R^2} \ln \Big(\frac{\rho_ \alpha}{R}\Big)\Big] (\hat{z} \times \xv_\alpha)_i ,
\end{multline}
where $\sigma_{ij}$ is the stress in the $\Dv= 0$ state.  Hence, in a bundle with a single off-center, 5-fold defect, the elastic energy decreases most rapidly by tilt in a direction $\pi/2$ away from the radial location of the dislocation.  When $\Dv \neq 0$, the center of rotation in the bundle is not the central filament, and eq. (\ref{eq: torque}) shows that this rotation center, for which $\tv_\perp(\xv) = 0$, migrates from the bundle center {\it towards} the off-center disclination by a distance, $|\Dv|/\Omega$, demonstrating that the low-energy state of bundle adopts a globally writhing and helical configuration. 

To investigate the buckled, helical groundstates of bundles with off-center disclinations, we introduce a discrete, numerical model of cylindrical bundles of fully flexible filments with preferred hexagonal order.  Bundles are described arrays of filaments at positions, $\xv_\alpha$, and filament tangents corresponding to, $\tv_\perp (\xv) = \Dv + \Omega (\hat{z} \times \xv) $.  The elastic energy in a vertical cross-section of height, $\Delta z$, is constructed following the bead-spring model of ref. \cite{nelson_seung},
\begin{equation}
\label{eq: num}
\Delta E= \frac{k}{2} \sum_{\langle  \alpha \beta \rangle} \big(|(\xv_\alpha-\xv_\beta)_\perp| - d_0\big)^2 \Delta \ell_{ \alpha \beta} .
\end{equation}
Here, $k$ parameterizes a harmonic restoring force for distortions in the packing $(\xv_\alpha-\xv_\beta)_\perp = \xv_\alpha- \xv_\beta - \tv(\xv_\alpha-\xv_\beta)\cdot \tv$ is the projection of the filament spacing {\it perpendicular} to the local tangent direction, thereby, measuring distances of closest approach for weakly bent filaments.  The interactions are weighted the {\it lengths} of elements $\Delta \ell_{\alpha \beta} = \Delta z (\tv \cdot \hat{z})^{-1}$ which vary with in-plane tilt in the cross-section.  To study the groundstates numerically, we generate a 5-fold disclination in an unrelaxed, infinite hexagonal array.  This disclination is offset to a distance, $\rho$, from the center of the bundle at a given azimuthal angle.  Lattice positions within a radial distance, $r \leq R = 10 d_0$, are triangulated and used as initial in-plane coordinates of filaments in the bundle.  The elastic energy per total filament length is numerically minimized by relaxing filament positions, $\Omega$ and $\Dv$.

\begin{figure}
\center \epsfig{file=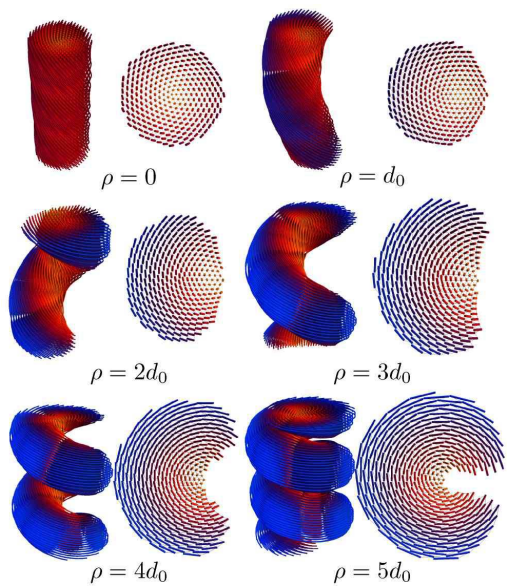, width=2.65in}\caption{Cross-sectional (constant $z$) and side views of energy minimizing bundles for the disclination positions, $\rho \leq 5 d_0$.  Filament colors correspond to radial distance from center of rotation in the $xy$ plane, from yellow $r =0$, to blue $r = 20 d_0$.  } 
\label{fig: figure3}
\end{figure}

The geometry of energy-minimizing configurations is plotted in Figure \ref{fig: figure2} in terms of the rate of torsion, $\Omega$, and the helical radius of the geometric center of the bundle, $r_h$, both of which reveal a striking sensitivity of global bundle geometry of the bundle to disclination position.  The center of helical rotation migrates rapidly from $r_h=0$ when $\rho=0$ to position nearly at the boundary of bundle, $r_h \simeq 9.7d_0$ when $\rho=5 d_0$, adopting a globally helical conformation.  Perhaps more surprising, the degree of spontaneous rotation, $\Omega$, steeply increases as the disclination position is offset from the bundle center.  Indeed, the helical pitch of the bundle decreases more than 3-fold as the disclination is moved from $\rho=0$ to $\rho=5 d_0$.  Remarkably, the writhing and contorted bundle geometries shown in Fig. \ref{fig: figure3} provide a more uniform filament spacing than can be achieved in straight bundles.  For disclinations at radial positions, $\rho \geq  6 d_0$, we find that bundle groundstates are {\it self-intersecting}, and therefore, cannot be modeled without the inclusion of either bend elasticity or excluded-volume constraints.  

In summary, we have demonstrated a coupling between non-linear elastic stresses that arise from twisting of a fillament bundle and orientational defects in the lattice packing.  Thus, 5-fold disclinations are necessary components of low-energy configurations of sufficiently twisted and large bundles of continuous filaments.  We expect this finding to be particularly relevant to the self-assembled bundles and fibers of biological filaments, whose chiral molecular structure leads to a generic preference for inter-filament twist in compact assemblies.  Indeed, cylindrical collagen fibrils derived from certain tissue types show pronounced twist, along with considerable evidence from small-angle scattering of localized disorder in their cross-sectional organization~\cite{wess, prockop}.     In light the results of this, we speculate that the unique properties of these self-organized collagen fibrils, the well-defined diameter, the twisted structure, and their ``almost crystalline" order, may all find a common origin in generic properties of the non-linear elasticity of filament assemblies.

\begin{acknowledgments}
It is a pleasure to acknowledge R. Meyer and M. Muthukumar for useful insights as well as P. Ziherl and A. Travesset for careful readings of this manuscript.  This work was supported by the NSF Career program under DMR Grant 09-55760.  The author acknowledges hospitality of the Aspen Center of Physics were some of this work was completed.

\end{acknowledgments}

\end{document}